\newif\ifsingle

\newif\ifFullVersion
\FullVersiontrue 

\ifsingle
\documentclass[11pt,draftclsnofoot, onecolumn]{IEEEtran}		
\else		
\documentclass[9pt,final, twocolumn]{IEEEtran}
\fi

\usepackage{times}
\usepackage{amsmath,dsfont}
\usepackage{amssymb,amsthm}
\usepackage{epsfig,verbatim}
\usepackage{subfigure}
\usepackage{setspace}
\usepackage{color}
\usepackage{cite}
\usepackage{epstopdf}
\usepackage{graphics}
\usepackage{accents}
\usepackage{acronym}
\usepackage[bookmarks,colorlinks]{hyperref}
\usepackage{booktabs}
\usepackage{mathtools}
\usepackage{algorithm}
\usepackage{algorithmic}
\usepackage{enumitem}

\newcommand{\mySet}[1]{\mathcal{#1}}

\newtheorem{proposition}{Proposition}

\ifsingle

\else

\fi 

\acrodef{dma}[DMAs]{dynamic metasurface antennas}
\acrodef{snr}[SNR]{signal-to-noise ratio}
\acrodef{sinr}[SINR]{signal-to-interference-and-noise ratio}
\acrodef{bs}[BS]{base station} 
\acrodef{em}[EM]{electromagnetic} 
\acrodef{mimo}[MIMO]{multiple-input multiple-output}
\acrodef{mmw}[mmWave]{millimeter wave}

\IEEEoverridecommandlockouts

\title{Joint Microstrip Selection and Beamforming Design for MmWave Systems with Dynamic Metasurface Antennas}
\author{  
	\IEEEauthorblockN{Wei Huang, Haiyang Zhang, Nir Shlezinger, and Yonina C. Eldar \\
	} 
\thanks{
    W. Huang is with the School of Computing Science and Information Engineering, Hefei University of Technology, Hefei, China (e-mail: huangwei@hfut.edu.cn). 
  H. Zhang is with School of Communications and Information Engineering, Nanjing University of Posts and Telecommunications, Nanjing, China (e-mail: 20220142@njupt.edu.cn). 
  N. Shlezinger is with the School of ECE, Ben-Gurion University of the Negev, Beer-Sheva, Israel (e-mail: nirshl@bgu.ac.il). 
  Y. C. Eldar is with the Faculty of Math and CS, Weizmann Institute of Science, Rehovot, Israel (e-mail: yonina.eldar@weizmann.ac.il).
	}
	\vspace{-1.0cm}
}
\vspace{-0.75cm}

\begin{document}
	
	\maketitle
	\pagestyle{empty}
	\thispagestyle{empty}
\begin{abstract}
Dynamic metasurface antennas (DMAs) provide a new paradigm to realize large-scale antenna arrays for future wireless systems. In this paper, we study the downlink millimeter wave (mmWave) DMA systems with limited number of radio frequency (RF) chains. By using a specific DMA structure, an equivalent mmWave channel model is first explicitly characterized. Based on that, we propose an effective joint microstrip selection and beamforming scheme to accommodate for the limited number of RF chains. A low-complexity digital beamforming solution with channel gain-based microstrip selection is developed, while the analog beamformer is obtained via a coordinate ascent method. The proposed scheme is numerically shown to approach the performance of DMAs without RF chain reduction, verifying the effectiveness of the proposed schemes.

\end{abstract}

\acresetall

\vspace{-0.2cm}
\section{Introduction}
In order to increase the capacity of  wireless communication systems, millimeter wave (mmWave) bands ranging from $30\mspace{2mu}\rm GHz$ to $300\mspace{2mu}\rm GHz$, are regarded as a promising candidate. The small wavelength of mmWave signals allows a large number of antenna elements to be be packed in a small area, facilitating multiple-input multiple-output (MIMO) processing with very large arrays. However, realizing large antenna arrays for mmWave communication systems in practice can be challenging. A key difficulty stems from the fact that radio frequency (RF) chains in mmWave are costly in terms of hardware implementation, signal processing complexity, and energy consumption. To overcome this issue, various cost-aware hybrid architectures with limited number of RF chains have been proposed \cite{7400949,8425998,gong2019rf}. However, typical hybrid architectures come at the cost of additional analog circuitry, typically comprised of multiple  phase shifters, which can lead to relatively high energy consumption~\cite{zirtiloglu2022power}. 

An alternative large-scale MIMO technology utilizes \ac{dma}~\cite{2008788}. \ac{dma} consist of multiple waveguides (microstrips) and each embedded with many metamaterial antenna elements, which inherently provides analog beamforming capabilities with lower power consumption and cost compared with typical phased array antennas~\cite{williams2022electromagnetic}. Moreover, \ac{dma}
are typically utilized with sub-wavelength element spacing, allowing to pack a larger number of elements in a given
antenna area compared to conventional phased array antennas. This ability has been exploited to improve communication capacity and energy efficiency  \cite{9324910, you2022energy}. 

Over the last few years, several transmission schemes with \ac{dma} have been studied under various wireless scenarios, such as orthogonal frequency multiplex systems \cite{9272351}, hybrid reconfigurable intelligent surfaces and \ac{dma} network \cite{9827859}, and wireless power transfer \cite{9833917}. The wireless channels considered in the above works are based on statistical models commonly employed in lower bands, i.e., Rayleigh or Rician distributions. However, for the high-frequency mmWave bandwidth, these models may be not applicable, and one should account for the geometry of the physical channel \cite{6717211}. In this case, the specific propagation characteristics of \ac{dma}, where signals at different elements propagate differently inside the waveguides \cite{8756024} is translated into an important consideration for transmission designs in mmWave communication systems. In particular, the equivalent physical channel depends on both the specific DMA structure and mmWave wireless channel model. On the other hand, due to the mmWave channel sparsity, only a small number of beams are needed. Since each beam corresponds to a single RF chain~\cite{7953407}, one can use only part of the \ac{dma} microstrips via a switching network, which can further reduce cost and power.

Based on the above observations, in this paper we propose a  joint microstrip selection and beamforming design scheme for mmWave signalling with \ac{dma}. We first explicitly characterize the equivalent mmWave physical channel model with a DAM, accounting for  the propagation characteristics of mmWave signals and the DAM structure. Based on the obtained equivalent channel model, we then study the hybrid beamforming design problem to maximize the single-noise ratio (SNR), subject to a limit on the number of activated RF chains. The formulated joint microstrip selection and beamforming design problem is non-convex due to both the non-convex $l_0$ norm constraint and the non-convex Lorentz constraint~\cite{zhang2022beam}. To overcome the issue, we develop an iterative algorithm to set the digital beamforming and configure the \ac{dma} weights, alternately. Our optimizer uses a channel-gain based microstrip selection and beamforming, while configuring the \ac{dma} weights  via the coordinate ascent method. Simulations result indicate that the proposed scheme could reduce the number of RF chains by $75\%$, while the spectral efficiency approaches the case without the RF chain constraint. 


\emph{Notations}: In this paper, the upper and lower case bold
symbols denote matrices and vectors, respectively. We use $({\cdot})^{\rm T}$, $(\cdot)^*$, $({\cdot})^{\rm H}$, $|\cdot|$, and $\|\cdot\|_p$ to denote the transpose, conjugate, Hermitian transpose, absolute value, and $p$-norm, respectively. ${\mathbb C}^{M\times N}$ is the space of $M \times N$ complex-value matrices, symbol $\angle(\cdot)$ denotes the angle, while $\otimes$ and $\odot$ denote the Kronecker and Hadamard product, respectively.

	\section{System Model and Problem Formulation}
	\label{sec:Model}
 \subsection{Dynamic Metasurface Antennas} \label{sub:DMA}
Here, we give a brief review of \ac{dma}. \ac{dma} are metasurface-based antennas comprised of multiple microstrips, which are one-dimensional arrays of metamaterial elements placed on a waveguide cavity~\cite{2008788}. In such architectures, each RF chain is connected with the port located at the edge of the microstrip. Thus, during transmission, the signals in the microstrip propagating from that port undergo a different path for each element, which results in different propagation delay depending on their location. Define $\alpha_n$ as the wavenumber along microstrip $n$, and $\rho_{n,m}$ denotes the relative location of the $m$th element of the $n$th microstrip, which is usually proportional to the distance between the port of microstrip $n$ and the $m$th element. Thus, element-dependent propagation effect is formulated as
\begin{align}\label{inside response}
f_{n,m}=e^{-\rho_{n,m}(\beta_m+j\alpha_n)}\,,\forall n,m\,,
\end{align}
where $\beta_m$ is the waveguide attenuation coefficient of element $m$.

Each metamaterial elements acts as a resonant circuit whose frequency response for  narrowband signaling is approximated as the Lorentzian-constrained phase weights model\cite{8756024,2017Analysis}, given by
\begin{align}\label{lorentzian response}
g_{n,m}\in {\cal G}\triangleq \left\{\frac{j+e^{j\phi_{n,m}}}2|\phi_{n,m} \in [0,2\pi]\right\}\,,\forall n,m\,.
\end{align}
Here, $\phi_{n,m}$ denotes the phase shift of the $n$th element in the $m$th microstrip. From \eqref{inside response} and \eqref{lorentzian response}, we observe that the equivalent \ac{dma} frequency response for each radiating element is a product of the propogation response \eqref{inside response}, which is dictated by the placing of the elements,  and the tunable Lorentzian phase weight~ \eqref{lorentzian response}.

\begin{figure}
 \centering
 \includegraphics[width=3in]{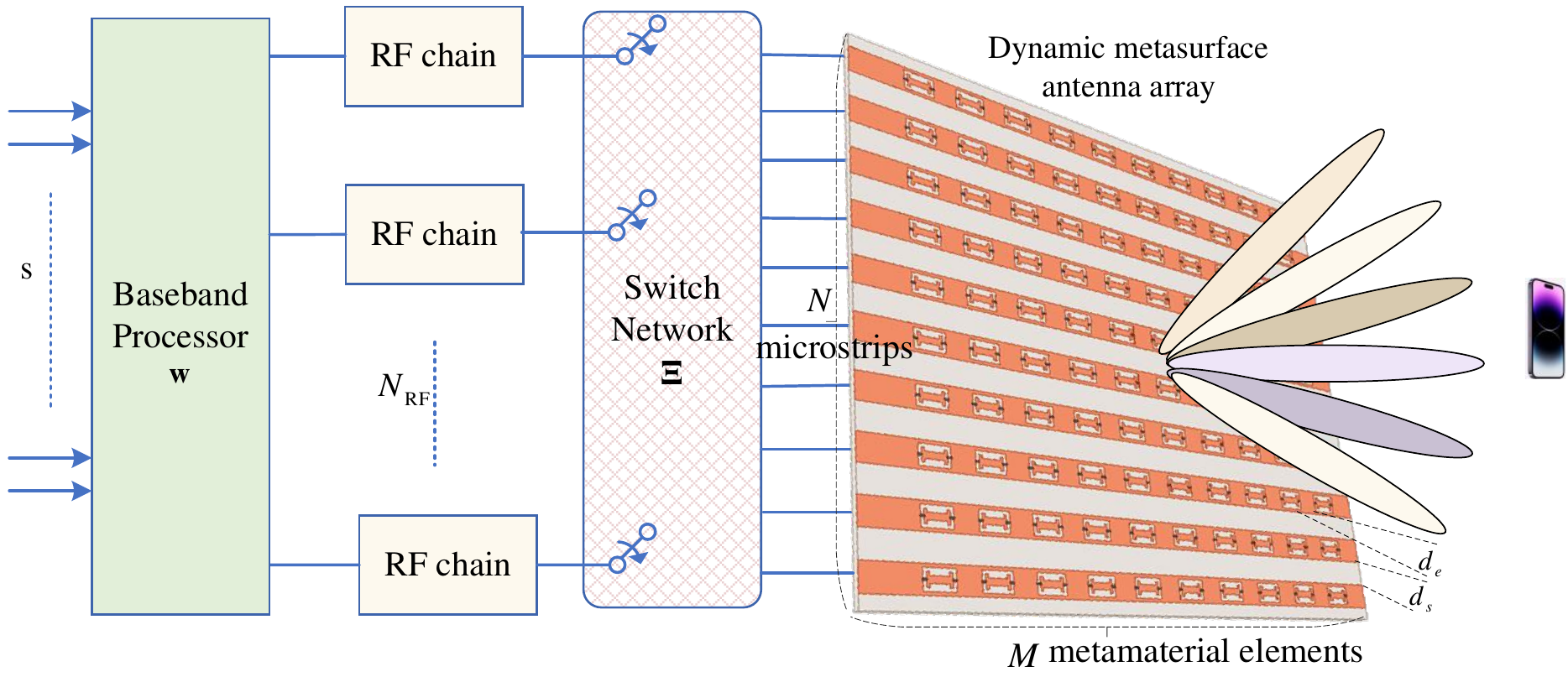} \caption{Downlink MISO system with DMAs.}     \label{fig1}
 \end{figure}

 
 
 \subsection{Downlink MmWave Systems with DMAs} \label{sub:model}
 We consider a mmWave downlink multiple-input single output (MISO) system as illustrated in Fig.~\ref{fig1}. Here, a base station (BS) equipped with a DMA having $U=NM$ radiating metamaterial elements serves a single-antenna user, where $N$ and $M$ denote the number of microstrips and radiating elements in each microstrip, respectively. For cost-effective implementations, the BS is equipped with $N_{\rm RF}$ ($N_{\rm RF} \leq N$) RF chains, which are connected with the DMA array via a switch network. We adopt the widely utilized geometric block-fading channel model, where the channel from the BS to user, denoted by $\hat{\bf  h}\in {\mathbb C}^{U \times 1}$, is given by 
\begin{align}
\label{eqn:ChannelModel}
\hat{\bf  h}=[\hat{\bf h}_1^{\rm H},\hat{\bf h}_2^{\rm H},\cdots,\hat{\bf h}_n^{\rm H}\cdots,\hat{\bf  h}_N^{\rm H}]^{\rm H}=\sum\nolimits_{l=1}^L\eta_l{\bf a}(\theta_l)\,.
\end{align}
In \eqref{eqn:ChannelModel}, ${\bf\hat h}_n\in {\mathbb C}^{M\times 1}$ denotes the channel from the $n$th microstrip to user, and $L$ is the number of scatters (paths) from the BS to the user, which is generally much smaller than the number of antennas because of the severe path loss in mmWave band; $\eta_l$ denotes the complex-value gain of the $l$th path; $\theta_l\in [0, \pi)$ represents the angle of deviate (AoD) of the $l$th path between the BS and user; while ${\bf a}(\theta_l)$ is the antenna array response vector corresponding to AoD $\theta_l$ between BS and user, which is given by
\begin{align}
{\bf a}(\theta_l)&\!=\!\left[{1, \cdots\! ,{e^{ - j\frac{{2\pi (M \!-\! 1){d_e}\sin (\theta_l)}}{\lambda}}}} \right]^{\rm T}
\!\!\otimes \!
\left[{1, \cdots\! ,{e^{-j\frac{{2\pi (N\!-\!1){d_s}\cos (\theta_l)}}{\lambda }}}} \right]^{\rm T}\notag\\
&=\left[{1, \cdots ,{e^{ - j\frac{{2\pi }}\lambda\Omega_u(\theta_l)}}},\cdots, {e^{ - j\frac{{2\pi }}\lambda\Omega_U(\theta_l)}}\right]^{\rm T}\in {\mathbb C}^{U}\,,
\end{align}
where $\lambda$ is the signal wavelength, $d_e$ is the distance between adjacent elements, and $d_s$ denotes the distance between the microstrips. $\Omega_{u=m\cdot N + n}(\theta_l) \triangleq md_e\sin(\theta_l)+nd_s\cos(\theta_l)$ can be interpreted as the spatial frequency of the $u$th  element corresponding to AoD $\theta_l$. 

\begin{figure}
 \centering
 \includegraphics[width=3in]{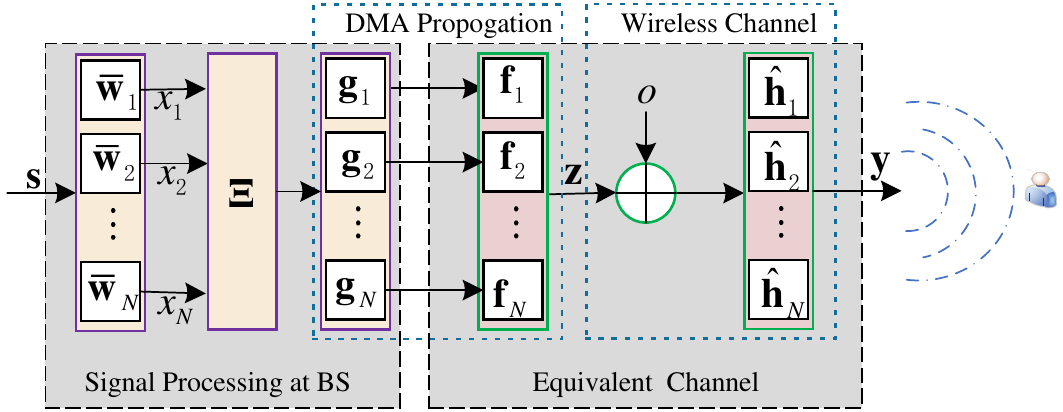} \caption{Illustration of the signal processing procedure.}     \label{fig2}
 \end{figure}

Since the outgoing  signals propagate inside the microstrips before being radiated, the equivalent MISO channel can be represented as
\begin{align}
{\bf h}=\left[{\bf h}_1^{\rm H},\cdots,{\bf h}_N^{\rm H}\right]^{\rm H}=\hat{\bf  h}\odot{\bf f}=\left(\sum\limits_{l=1}^{L}\eta_l{\bf a}(\theta_l)\right) \odot {\bf f}\,,
\end{align}
with
\begin{align}
{\bf f}
&=\left[f_{1,1},f_{1,2},\cdots,f_{1,M},\cdots,f_{n,m},\cdots,f_{N,M}\right]^{\rm T}\in {\mathbb C}^{U}\,.
\label{eqn:FDef}
\end{align}
The notation $f_{n,m}$ encapsulates the phase due to the distance propagated by the wave from the port of microstrip to the $n$th element. The vector $\bf f$ can be interpreted as the array response vector of the \ac{dma}. An illustration of the effective channel is depicted in Fig.~\ref{fig2}. Differently from  conventional phased-array or fully digital architectures, where the propagation characteristic of the wireless channel is not changeable, the outgoing signal undergoes the combined effect of the wireless channel and the microstrips for \ac{dma}, due to the element-dependent propagation inside the microstrip.

For a narrowband mmWave transmitter, the input signal at each microstrip can be expressed as
\begin{align}
x_n=w_ns\,, \qquad \forall n\,,
\end{align}
where $x_n$ is the input signal for microstrip $n$, $w_n$ is the corresponding digital beamforming, and $s$ is the information-bearing symbol with normalized power. Then, the baseband signals $x_n$ are fed to the corresponding microstrips via the switch network. Thus, the signal emitting by the $m$th element in microstrip $n$ is $f_{n,m} z_{n,m}$, where
\begin{align}\label{transmit signal}
z_{n,m}&=g_{n,m}\xi_{n}x_n=g_{n,m}\xi_{n}w_ns\,,
\end{align}
where $\xi_{n} \in \{0,1\}$ denotes whether the $n$th microstrip is activated or not. Define the $N\times 1$ vectors ${\bf x}=[x_{1},\cdots,x_N]^{\rm H}$,  ${\bf w}=[\xi_{1} w_1,\cdots, \xi_{N} w_N]^{\rm H}$. 
Then, \eqref{transmit signal} can be compactly written as
\begin{align}
{\bf z}={\bf  G}{\bf w}s\,,
\end{align}
where ${\bf G}={\rm diag}([{\bf g}_1^{\rm H},\cdots,{\bf g}_N^{\rm H}])\in {\mathbb C}^{U\times U} $ with  ${\bf g}_n=[g_{n,1},\cdots,g_{n,M}]^{\rm H}$ representing the configurable weights of microstrip $n$. After propagating via the wireless channel, the received signal $y$ is expressed as
\begin{align}
y={\bf h}^{\rm H}{\bf z}+o={\bf h}^{\rm H}{\bf  G}{\bf w}s+o\,,
\end{align}
where $o$ is  additive white Gaussian noise with variance $\sigma^2$. Then, the achievable rate is given by
$
{\rm log}\left(1+\frac{1}{\sigma ^2}|{{\bf h}^{\rm H}}{\bf{G w}}{|^2}\right)$.

 \subsection{Problem Formulation} \label{sub:problem}
 \vspace{-0.1cm}
 For a mmWave channel with severe path loss, a large-scalar array can be exploited to compensate for the high path loss. Even for a single antenna user, the number of antennas in one microstrip may not be sufficient, that is the fact the emitting energy of back-end antennas will become weaken when the signals are propagating inside the microstrip. It is thus of importance to properly utilize the available RF chains to achieve reliable communications. 
 
 Since the number of RF chains at the BS is generally smaller than that of the microstrips in mmWave systems, i.e., $N_{\rm RF}\le N$, the RF chains have to selectively connect with some microstrips via the switch network. As such, the number of active microstrips can not exceed $N_{\rm RF}$. Moreover, as the achievable rate monotonically grows with the SNR,  our objective is to maximize the SNR under the RF chains constraint, by joint microstrip selection and beamforming optimization. Therefore, the optimization problem is stated as 
 \begin{subequations}\label{single-user sparse}
\begin{align}
\mathop {\max }\limits_{{\bf w},\{g_{m,n}\}\in {\cal G}}\quad&\frac{1}{\sigma^2}|{\bf h}^{\rm H}{\bf Gw}|^2\\
{\rm s.t.}\quad&\|{\bf w}\|_2^2\le P\\
&\|{\bf w}\|_0 \le N_{\rm RF}.
\end{align}
\end{subequations}
Constraint (\ref{single-user sparse}c) guarantees that at most $N_{\rm RF}$ microstrips are activated, while  (\ref{single-user sparse}b) represents the power constraint. 
Note that problem \eqref{single-user sparse} is  non-convex due to the non-convex constraints (\ref{single-user sparse}c) and the coupling of optimization variables $\bf w$ and {\bf G} in objective function (\ref{single-user sparse}a). Furthermore, different from phased array based hybrid architecture, the analog beamforming of \ac{dma} is subjected to the Lorentz constraint, i.e. $\{g_{n,m}\}\in \cal G$. To solve the non-convex problem, we develop an iterative algorithm in following section.

	\section{Joint Microstrip and Beamforming Design}
	\label{sec:Solution}
In this section, we develop an  algorithm to optimize $\bf w$ and the  weights $g_{m,n}$. We adopt an alternating approach, where we maximize the SNR by sequentially fixing one variable and updating the other. 	

{\bf Optimizing ${\bf w}$:}
For a given ${\bf G}$, the goal is to
jointly select the  $N_{\rm RF}$ out of $N$ microstrips and design the
corresponding beamforming vector associated with the selected
microstrips to maximize the SNR. The resultant subproblem is given by
\begin{subequations}\label{digital formu}
\begin{align}
{\mathcal P}_1:\mspace{5mu}\mathop {\max }\limits_{{\bf w}}\quad&\frac{1}{\sigma^2}|{\bf h}^{\rm H}{\bf Gw}|^2\\
{\rm s.t.}\quad&\|{\bf w}\|_2^2\le P\\
&\|{\bf w}\|_0 \le N_{\rm RF}\,.
\end{align}
\end{subequations}

Note that in ${\cal P}_1$, the objective  (\ref{digital formu}a) and power constraint (\ref{digital formu}b) are convex. The main difficulty lies in the non-convex and non-differentiable cardinality constraint (\ref{digital formu}c), which restricts the number of the selected microstrips to be no more than $N_{\rm RF}$. Despite this, ${\cal P}_1$ admits a closed-form solution, as stated in the following:
\begin{proposition}
\label{pro:OptW}
Let $\mySet{I} \triangleq \{i_1,\ldots,i_{N_{\rm RF}}\}$ denote the indices of the $N_{\rm RF}$ largest entries in the set $\{|{\bf h}_n^{\rm H} {\bf g}_n|^2\}_{n=1}^{N}$. Accordingly, define ${\bar {\bf h}}  \triangleq [{\bf h}_{i_1},\ldots, {\bf h}_{i_{N_{\rm RF}}}]$,  ${\bar {\bf G}}\triangleq {\rm diag}([{\bf g}_{i_1}^{\rm H},\cdots,{\bf g}_{i_{N_{\rm RF}}}^{\rm H}])$, and
\begin{align}\label{opti trans full}
{\bar{\bf w}}=\sqrt P\frac{{\bar {\bf G}}^{\rm H}{\bar {\bf h}}}{\|{\bar {\bf G}}^{\rm H}{\bar {\bf h}}\|}\,.
\end{align}
Then, it holds that ${\cal P}_1$ is solved by setting 
\begin{equation}
\label{eqn:OptW}
    w_n = \begin{cases}
    [{\bar{\bf w}}]_j & \exists i_j \in \mySet{I}  \text{ such that } n=i_j \\
    0 & \text{otherwise.}
    \end{cases}
\end{equation}
\end{proposition}
\ifFullVersion
\begin{IEEEproof}
The proof is given in Appendix \ref{app:Proof1}.
\end{IEEEproof}
\fi
\smallskip
{\bf Optimizing ${\bf G}$:}
For a given beamforming vector ${\bar {\bf w}}$ set via Proposition~\ref{pro:OptW}, the resulting SNR is given by
\begin{align}
{\rm SNR} &= \frac{1}{\sigma^2}|{\bar{\bf h}}^{\rm H}{\bar{\bf G}}{\bar {\bf w}}|^2 
=\frac{P}{\sigma^2}\|{\bar {\bf h}}^{\rm H}{\bar {\bf G}}\|^2 \notag \\ &=\frac{P}{\sigma^2}\sum\nolimits_{n=1}^{N_{\rm RF}}\left|{\sum\nolimits_{m=1}^M  {h}^*_{i_n,m} {g}_{i_n,m}}\right|^2\,,
\label{eqn:SNRExpG}
\end{align}
where $h_{i_n,m}$ denotes the $m$th element of vector ${
\bf h}_{i_n,m}$. Then, the configurable weights optimization subproblem is posed as
\begin{align}\label{confi opti}
{\cal P}_2:\mspace{5mu}\mathop {\max}\limits_{\{ g_{i_n,m}\} \in {\cal G}}\quad&\frac{P}{\sigma^2}\sum\nolimits_{n=1}^{N_{\rm RF}}\left|{\sum\nolimits_{m=1}^M{ h}_{i_n,m}^*{ g}_{i_n,m}}\right|^2.
\end{align}
Note that in \eqref{confi opti}, the Lorentzian constraint in ${\cal P}_2$ characterizes the
feasible set as a circle on the complex plane $\left|g-\frac12e^{j\frac\pi 2}\right|$ $=\frac 1 2$, with the circle center at $(0,e^{j\frac \pi 2})$ and radius equals to $\frac 1 2$. This non-convex constraint makes the subproblem ${\cal P}_2$ difficult to be solved directly. To address the this, we define the variables $b_{n,m},\forall m,n$, which are related to the DMA weights via the affine mapping
\begin{align}\label{replace variable}
b_{n,m}=2{g}_{i_n,m}-e^{j\frac\pi 2}\,,\forall m,n\,.
\end{align}
The variable $b_{n,m}, \forall m,n$ lies on the unit circle of complex plane with the circle center at the origin of coordinates. With \eqref{replace variable}, the Lorentzian constraint in ${\cal P}_2$ is converted to the unit-amplitude constraint. Thus, subproblem ${\cal P}_2$ is equivalently transformed into
\begin{align}\label{convert form}
{\cal P}_{2.1}:\mspace{1mu}\mathop{\max}\limits_{\{b_{n,m}\}}\mspace{3mu}&\frac P {4\sigma^2} \sum\nolimits_{n=1}^{N_{\rm RF}}\left|\sum\nolimits_{m=1}^M\left({ h}_{i_n,m}^*b_{i_n,m}+{ h}_{i_n,m}^*e^{j\frac \pi 2}\right)\right|^2\notag\\
{\rm s.t.}\mspace{3mu} & |b_{n,m}|^2=1\,,\mspace{5mu}\forall m,n\,.
\end{align}
While ${\cal P}_{2.1}$ is still challenging, it can be tackled via   a coordinate ascent-based heuristic algorithm to find a local optimal solution, based on the following proposition.
\begin{proposition}
\label{Lem:OptimalPhase}
Any local optimal phase of \eqref{convert form} satisfies
\begin{align}\label{lemma 1}
\angle b_{n,m}= - \angle \left(h^*_{i_n,m}{\tilde h}^*_{n,m}\right)\,\quad\forall n,m\,,
\end{align}
where 
${\tilde h}_{n,m}=\sum\limits_{m'\neq m}  h^*_{i_n,m}b_{n,m'}+\sum\nolimits_{m=1}^M h^*_{i_n,m}e^{j\frac \pi 2}$.
\end{proposition}

\ifFullVersion
\begin{IEEEproof}
The proof is given in Appendix \ref{app:Proof2}.
\end{IEEEproof}
\fi

\smallskip
Using Proposition~\ref{Lem:OptimalPhase}, we can find the local optimal solution of subproblem ${\cal P}_{2.1}$ by alternately updating each element $b_{n,m}$, $n \in \mathcal{I}$, with all other elements fixed. Since the objective function in ${\cal P}_{2.1}$ always increases in each iteration, the iteration algorithm can guarantee to converge eventually. Then, based on the affine relationship between $g_{i_n,m}$ and $b_{n,m}$  in \eqref{replace variable}, we configure the weights of the elements in the active microstrips $\{g_{n,m}\}_{n \in \mathcal{I}}$.

\smallskip
{\bf Algorithm Summary:} 
The joint optimization of the \ac{dma} weights along with the microstrip selection and beamforming alternates between the individual settings in Propositions~\ref{pro:OptW}-\ref{Lem:OptimalPhase}. 
In each iteration, we take the solutions $\bar {\bf w}$ and $\{ g_{i_n,m}\}$ into those locations corresponding to the selected mircostrips of $\bf w$ and $\bf G$, and set the entries that are non-selected microstrips zeros, respectively. The proposed algorithm to solve problem \eqref{single-user sparse} is summarized as Algorithm~\ref{alg1}.
\begin{algorithm}
\caption{Solving problem \eqref{single-user sparse} w.r.t. $\bf w$ and $\{g_{n,m}\}$}
\label{alg1}
\begin{algorithmic}[1]
\STATE  {\bf Initialize:} set intial {\bf G} and threshold $\epsilon>0$  \\
\REPEAT
\STATE  Get $\mathcal{I}$ and ${\bf w}$ based on ${\bf G}$ via Proposition~\ref{pro:OptW};\\
\STATE  Set $\{g_{n,m}\}_{n \in \mathcal{I}}$ via \eqref{lemma 1} and \eqref{replace variable}
\UNTIL {SNR increase is smaller than $\epsilon$}
\STATE \textbf{Output:} Beampattern $\bf w$; active set $\mathcal{I}$, and DMA weights $\{g_{n,m}\}$
\end{algorithmic}
\end{algorithm}

{\bf Discussion:} 
The proposed Algorithm~\ref{alg1} jointly designs the digital beamformers, DMA weights, and microstrip selection, via alternating optimization. Each step in Algorithm~\ref{alg1} is based on a simple closed-form computation, and it is particularly designed to be simple to implement such that it can be carried out online. The resulting joint design allows to achieve a rate within a relatively small gap from that of costly fully-digital arrays, as we demonstrate in Section~\ref{sec:Numerical}. 

Our joint design gives rise to multiple potential extensions.
The fact that the signal processing capabilities of DMAs can be viewed as a form of hybrid precoding indicates that the proposed Algorithm can be extended to accommodate other forms of hybrid architectures based on, e.g., vector modulators \cite{zirtiloglu2022power} and phase shifter networks \cite{ioushua2019family}. Furthermore, while we focus here on narrowband signalling, for wideband signals, \ac{dma} are known to provide additional degrees of freedom in the form of a flexible frequency selective analog processing \cite{9324910}. Moreover, when operating in rapidly time-varying channels, one may wish to limit the number of alternating iterations to a small number. In such cases, emerging model-based deep learning techniques \cite{shlezinger2022model} can leverage data to facilitate rapid processing, see, e.g., \cite{agiv2022learn}. We leave these extensions for future study.



\begin{figure}
 \centering
 \includegraphics[width=0.8\columnwidth]{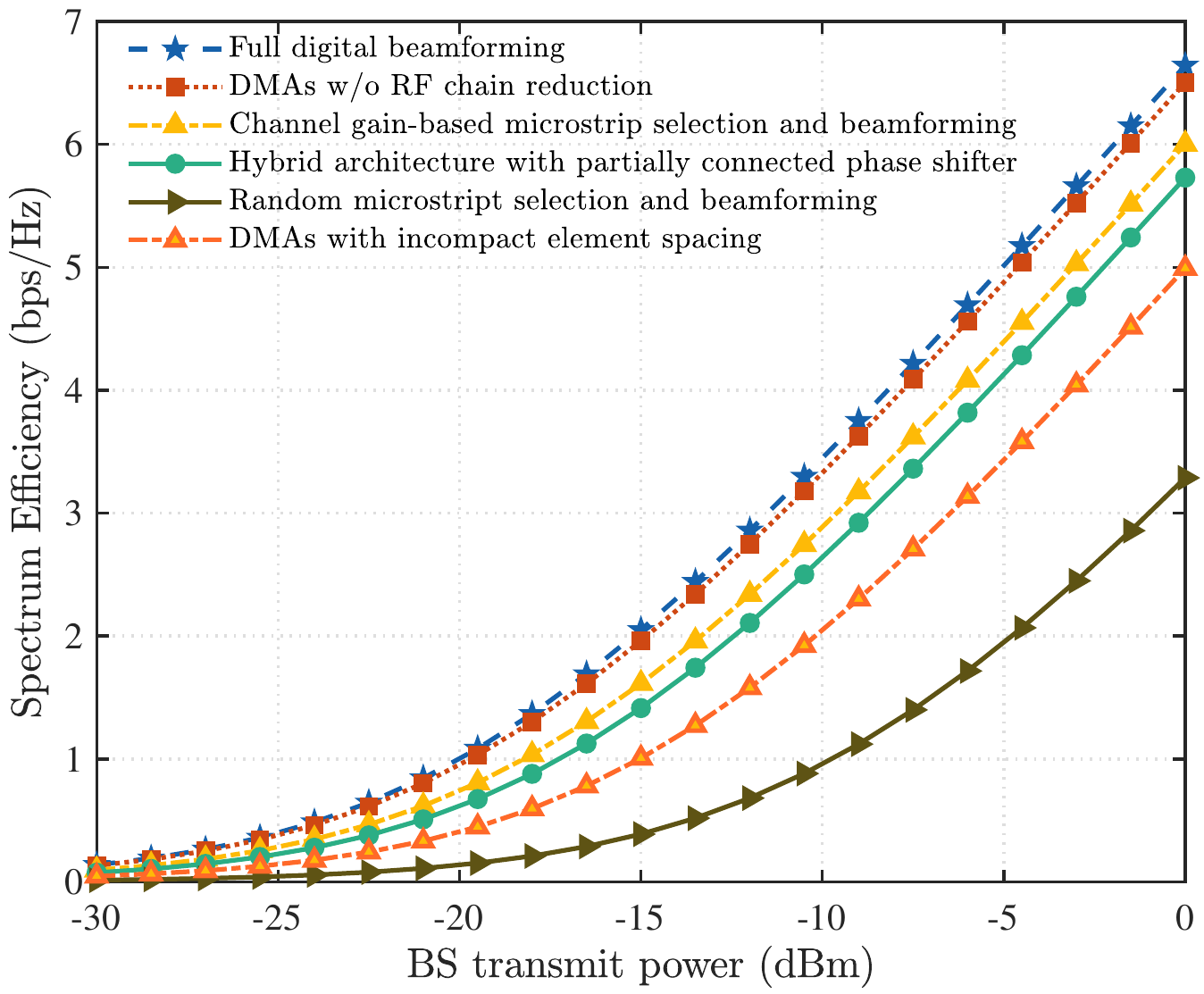} \caption{ Spectrum efficiency versus BS transmit power.}     \label{single_SNR}
 \end{figure}

	\vspace{-0.2cm}
	\section{Numerical Evaluations}
	\label{sec:Numerical}
	\vspace{-0.1cm} 
In this section, we numerically evaluate our proposed design. Unless otherwise stated, we consider a planar array located in the $xz-$plane. 
All simulated architectures have the same-sized array aperture and the array size is $0.3\mspace{2mu}{\rm m}\times 0.3\mspace{2mu}{\rm m}$. For \ac{dma}, $\lambda/5$ spacing  between the elements inside each microstrip is considered and the separation between microstrips could be still $\lambda/2$. That is, $N=10$ and $M=30$. While for the full digital architecture and hybrid architecture with partially connected phase shifter, the element spacing is $\lambda/2$, i.e., $N=M=10$.  The system operates at $28 {\rm GHz}$ and the mmWave channel has $12$ clusters. As in \cite{zhang2022beam}, we set $\alpha=0.6[\rm m^{-1}]$ and $\beta=827.67[\rm m^{-1}]$ to represent the propagation inside the DMA waveguides, representing a microstip implemented in Duroid $5880$. 

Our proposed design is compared with the following schemes:\\
$\quad~\bullet$  \emph{Fully digital beaforming}: Each antenna is connected to a dedicated RF chain and the corresponding transmit beamforming uses the maximum ratio transmit (MRT) strategy~\cite{5657271}.\\
$\quad~\bullet$ \emph{\ac{dma} architecture without RF chain reduction}:  Each microstrip is connected with one RF chain, and the digital and analog beamforming are designed as in \cite{zhang2022beam}. \\
$\quad~\bullet$ \emph{\ac{dma} architecture with incompact element spacing}: All elements spacing are set to $\lambda/2$, and the digital and analog beamforming are designed using the approach proposed in Section \ref{sec:Solution}.\\
$\quad~\bullet$  \emph{Random microstrip selection and beamforming}: Randomly select the given number of microstrips, and then apply the proposed method in \ref{sec:Solution} to design the corresponding digital and analog beamforming.
\\
$\quad~\bullet$  \emph{Hybrid architecture with partially connected phase shifter}: The number of connected RF chains is same to the random microstrip selection scheme, and the digital and analog beamforming vectors are derived via the alternate optimization approach proposed in \cite{8457260}.

In Fig.~\ref{single_SNR}, we compare the spectrum efficiency of our proposed Algorithm~\ref{alg1} (channel gain-based microstrip selection and beamforming) with the above-mentioned benchmark schemes. For microstrip reduction schemes, the total number of RF chains is $N_{\rm RF}=3$. It is observed that the performance proposed in Algorithm~\ref{alg1} is better than that of the hybrid architecture with partially connected phase shifter. The gain is attributed to the sub-wavelength feature of \ac{dma}, allowing to pack a larger number of elements in a given
physical antenna area compared to phased array antennas. However, for \ac{dma} with incompact element space, the performance of hybrid architecture with partially connected phase shifter is better than that of \ac{dma}. 
Moreover, we can see that the achievable spectrum efficiency of our proposed Algorithm~\ref{alg1} is comparable to that of DMA architecture without RF chain reduction. That is to say, our proposed scheme is capable of reducing the number of RF chains by $75\%$ with negligible performance degradation.

\begin{figure}
 \centering
 \includegraphics[width=0.8\columnwidth]{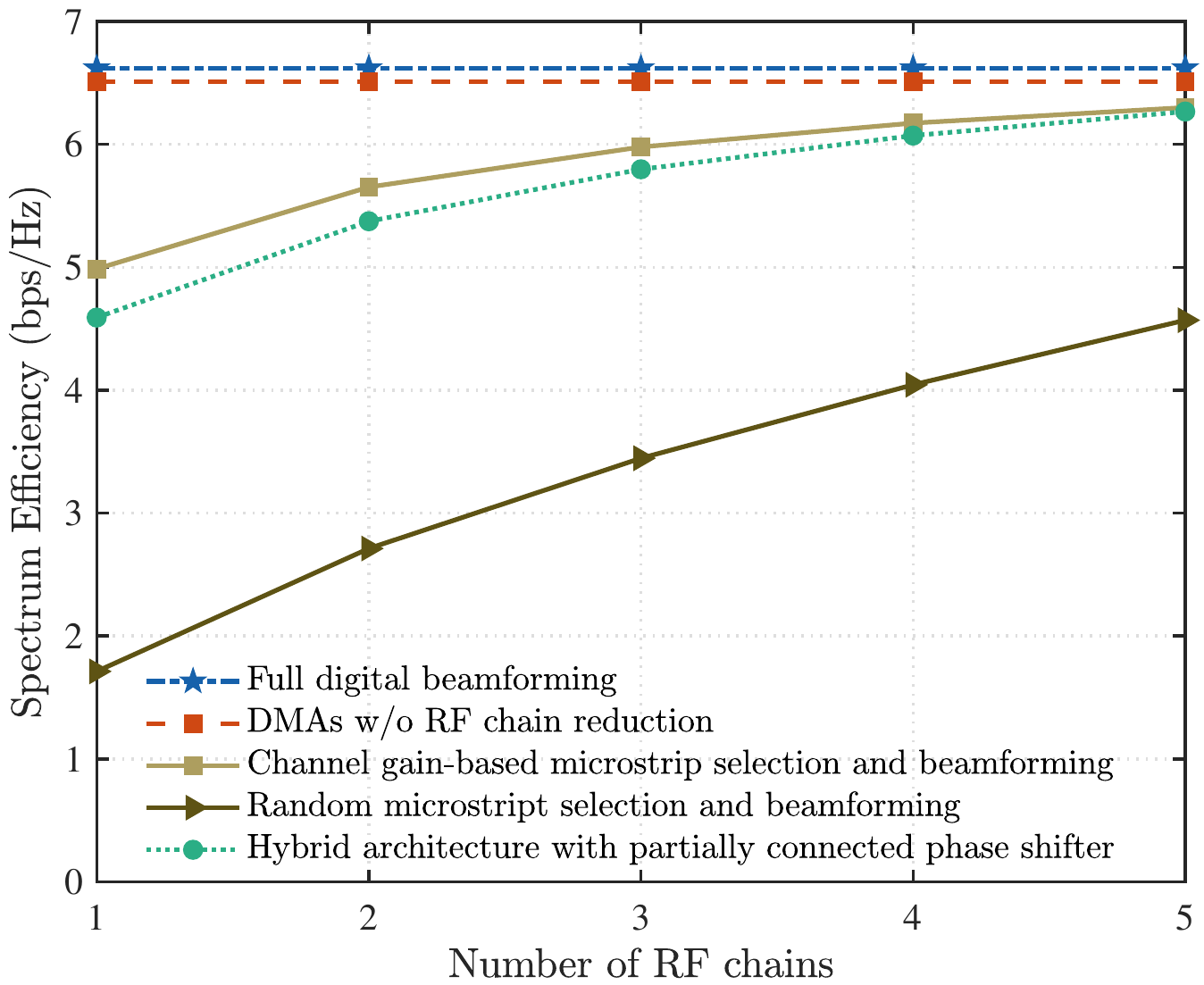} \caption{ Spectrum efficiency versus number of RF chains.}     \label{single_RF}
 \end{figure}

Fig.~\ref{single_RF} depicts spectrum efficiency versus the number of RF chains $N_{\rm RF}$ at the BS, where the transmit power is set as $0 \mspace{2mu}\rm dBm$. It indicates that for all schemes except fully digital and DMA without RF chains reduction, the performance in general improves with more RF chains, but only marginal improvement is observed as $N_{\rm RF}$ exceeds $5$. This is expected since performance will be constrained by the number of paths of mmWave channels and size of array aperture. Therefore, by using the proposed joint microstrip selection and beamforming scheme, the DAM can exploit the fewer RF chains to selectively connect to the microstrips so as to attain the performance with the fully RF chains, which can efficiently reduce the cost and power.

	\vspace{-0.2cm}
	\section{Conclusions}
	\label{sec:Conclusions}
	\vspace{-0.1cm}
In this paper, we studied mmWave MISO communications with DMA and limited number of RF chains. An equivalent mmWave channel model with DMA was characterised. Then, based on this model, a joint optimization problem of DMAs weights, RF chain selection matrix and digital beamforming vector is formulated to maximize the SNR. We  proposed an alternating optimization algorithm to solve the non-convex problem. Numerical results show that the proposed scheme can reduce the number of RF chains by $75\%$ and signal processing complexity without compromising on performance.

\ifFullVersion
\vspace{-0.2cm}
\begin{appendix}
	\numberwithin{proposition}{subsection} 
	\numberwithin{lemma}{subsection} 
	\numberwithin{corollary}{subsection} 
	\numberwithin{remark}{subsection} 
	\numberwithin{equation}{subsection}	
	%
	\vspace{-0.2cm}
	\subsection{Proof of Proposition \ref{pro:OptW}:}
	\label{app:Proof1}	
The SNR expression can be rewritten as
\begin{align}\label{convert snr}
   |{\bf h}^{\rm H}{\bf Gw}|^2=\sum\limits_{n=1}^N|{\bf h}_n^{\rm H}{\bf g}_nw_n|^2 =\sum\limits_{n=1}^N|{\bf h}_n^{\rm H}{\bf g}_n|^2|w_n|^2\,.
\end{align}
Thus, to maximize the SNR with the limited number of RF chains ($N=N_{\rm RF}$), we just need to select $N_{\rm RF}$ largest entries in the set $\{|{\bf h}_n^{\rm H}{\bf g}_n|^2\}$ and find the corresponding channels $\bar {\bf h}$ and \ac{dma} weights ${\bar{\bf G}}$. As a result, subproblem ${\cal P}_1$ becomes
\begin{align}
{\cal P}_{1.1}:\mspace{5mu}\mathop {\max }\limits_{\bar{\bf w}}\quad\frac{1}{\sigma^2}|{\bar {\bf h}}^{\rm H}{\bar {\bf G}}{\bar {\bf w}}|^2\quad
{\rm s.t.}\quad\|{\bar{ \bf w}}\|_2^2\le P\,,
\end{align}
where ${\bar{\bf w}}\in{\mathbb C}^{N_{\rm RF}\times 1}$ is the dimension-reduced digital beamforming vector. The optimal solution can be obtained via the MRT, i.e.,
\begin{align}\label{opti_w}
{\bar{\bf w}}^\star=\sqrt P\frac{{\bar {\bf G}}^{\rm H}{\bar {\bf h}}}{\|{\bar {\bf G}}^{\rm H}{\bar {\bf h}}\|}\,.
\end{align}
Then, for the selected microstrips, the beamforming coefficients can be extracted from \eqref{opti_w}; otherwise, set it to zero.\qed

	%
	\vspace{-0.2cm}
	\subsection{Proof of Proposition \ref{Lem:OptimalPhase}}
	\label{app:Proof2}	
Considering the fact that $|b_{n,m}|^2=1$ and write it as $e^{j \theta_{n,m}}$, the objective function of \eqref{convert form} can be expanded as
\vspace{-0.1cm}
\begin{align}
&\sum\limits_{n=1}^{N_{\rm RF}}\left|\sum\limits_{m=1}^M(h_{i_n,m}^*e^{j \theta_{n,m}}+h^*_{i_n,m}e^{j\frac \pi 2})\right|^2=\sum\limits_{n=1}^{N_{\rm RF}}|h^*_{i_n,m}e^{j \theta_{n,m}}\notag\\
&+\underbrace {\sum\limits_{m'\ne m} {{h^*_{i_n,m}}{e^{j\theta_{n,m'}}}+ \sum\limits_{m=1}^M{ h^*_{i_n,m}}{e^{j\frac\pi 2}}}}_{{{\tilde h}_{n,m}}}|^2\notag\\
&=\sum\limits_{n=1}^{N_{\rm RF}}| h^*_{i_n,m}e^{j \theta_{n,m}}+{\tilde h}_{n,m}|^2=\sum\limits_{n=1}^{N_{\rm RF}}|h^*_{i_n,m}|^2+|{\tilde h}_{n,m}|^2\notag\\
&+2{\rm Re}\{ h^*_{i_n,m}{{\tilde h}^*_{n,m}e^{j\theta_{n,m}}}\}\,.
\vspace{-0.1cm}
\end{align}
With all elements of variables $\{b_{n,m}\}$ fixed except $b_{n,m}$, we should maximize ${\rm Re}\{ h^*_{i_n,m}{{\tilde h}^*_{n,m}e^{j\theta_{n,m}}}\}$, which is equivalent to minimize the angle between $ h^*_{i_n,m}{\tilde h}^*_{n,m}$ and $e^{j\theta_{n,m}}$. Thus, we have
\begin{align*}
\theta_{n,m}^\star= - \angle \left( h^*_{i_n,m}{\tilde h}^*_{n,m}\right)\,\quad\forall n,m\,,
\end{align*}
which completes the proof.
\qed
	
\end{appendix}	
\fi 
 
\bibliographystyle{IEEEtran}
\bibliography{IEEEabrv,refs}

\end{document}